\documentstyle[12pt]{article}

\def \be{\begin{equation}}
\def \ee{\end{equation}}
\def \bea{\begin{eqnarray}}
\def \eea{\end{eqnarray}}

\begin{document}

\baselineskip 7mm

\title{Monte Carlo Hamiltonian}

\author{ H. Jirari, H. Kr\"{o}ger\thanks{Corresponding author: 
Tel: 1-418-656-2759,
Fax: 1-418-656-2040,
E-mail: hkroger@phy.ulaval.ca } \\ 
{\small\sl D\'epartement de Physique, Universit\'e Laval,
Qu\'ebec, Qu\'ebec G1K 7P4, Canada } \\
X.Q. Luo \\ 
{\small\sl CCAST(World Laboratory), P.O. Box 8730, Beijing 100080, 
China } \\
{\small\sl Department of Physics, Zhongshan University, Guangzhou 510275, 
China\thanks{Official and mailing address. 
E-mail: stslxq@zsulink.zsu.edu.cn} } \\
{\small\sl Center for Computational Physics, School of Physics Science and Engineering, } \\
{\small\sl Zhongshan University, Guangzhou 510275, China }\\
\and 
K.J.M. Moriarty\thanks{E-mail: moriarty@cs.dal.ca} \\
{\small\sl Department of Mathematics, Statistics and Computational Science, } \\
{\small\sl Dalhousie University, Halifax, Nova Scotia B3H 3J5, 
Canada } }


\maketitle

\begin{flushleft}
{\bf Abstract}
\end{flushleft}
We construct an effective Hamiltonian via Monte Carlo from a given action. This Hamiltonian describes physics in the low energy regime. We test it by 
computing spectrum, wave functions and thermodynamical observables (average energy and specific heat) for the free system and the harmonic oscillator. The method is shown to work also for other local potentials.

\begin{flushleft}
PACS index: o3.65.-w, 05.10.Ln 
\end{flushleft}

\setcounter{page}{0}

\newpage
\section{Motivation}\label{sec:Motivation}
\noindent The motivation for constructing a Monte Carlo Hamiltonian comes from different directions. \\
(i) The renormalization group \`a la Kadanoff-Wilson \cite{kn:Kadanoff} aims to construct a renormalized Hamiltonian, which describes physics at a critical point, based on the assumption of scale invariance. Such Hamiltonian is supposed to have much less degrees of freedom than the original Hamiltonian. A recent further development of those ideas is White's density matrix renormalization group technique \cite{kn:White}.
Our goal is similar to the above in the sense that we aim at an effective Hamiltonian, which has "less" degrees of freedom than the "original" Hamiltonian. But it differs in describing physics in the low energy domain instead of doing so at the critical point. \\
(ii) When one tries to solve field theory in the Hamiltonian formulation, the standard way to proceed is by constructing a Fock space, parametrized by some high momentum cut-off and some occupation number cut-off (Tamm-Dancoff approximation). When increasing those parameters, which means increasing the upper bound of the energy, then typically the density of states increases in an exponential manner, which renders the system beyond any control. In contrast to that, the Monte Carlo Hamiltonian is governed by a "small" number of low-energy degrees of freedom and the spectral density decreases with increasing energy. \\
(iii) The enormous success of lattice field theory over the last quarter of the century is certainly due to the fact that the Monte Carlo method with importance sampling is an excellent technique to solve high dimensional (and even "infinite" dimensional) integrals.
Conventionally, one computes a transition amplitude of an operator and evaluates it numerically via Monte Carlo (e.g. Metropolis algorithm \cite{kn:Metropolis}),
\bea
<O> &=& \frac{ \int [dx] ~ O[x] ~ \exp(- \frac{1}{\hbar} S[x]) }
             { \int [dx] ~ \exp(- \frac{1}{\hbar} S[x]) }
\nonumber \\
&\approx& \frac{1}{N_{c}} \sum_{C} O[C] .
\label{eq:PathIntegral}
\eea
Here $C$ stands for a path configuration drawn from the distribution 
$P[x] = \frac{1}{Z} \exp(-\frac{1}{\hbar} S[x])$.
The vitue of the Monte Carlo method lies in the property of yielding 
very good numerical results. E.g., solving a field theory model on a lattice of size $20^{4}$ and measuring the observable from a number of configurations 
$N_{c}$ in the order of a few hundred typically yields results with statistical errors in the order of a few percent. In this way it has been possible to determine low lying baryon and meson masses quite precisely \cite{kn:Weingarten}.

\bigskip

On the other hand, one can express a transition amplitude in imaginary time via the Hamiltonian
\bea
<x_{fi},T | x_{in},0  > &=& <x_{fi} | e^{-H T/\hbar} | x_{in}>
\nonumber \\
&=&
\sum_{n=1}^{\infty} < x_{fi} | E_{n}> e^{-E_{n} T/\hbar} 
< E_{n} | x_{in} >
\nonumber \\
&\approx& 
<x_{fi} | e^{-H_{ef} T/\hbar} | x_{in}>
\nonumber \\
&=& \sum_{\nu=1}^{N} < x_{fi} | E^{eff}_{\nu}> e^{-E^{eff}_{\nu} T/\hbar} 
< E^{eff}_{\nu} | x_{in} > .
\eea
In the last two lines we have approximated the Hamiltonian
$H$ by an effective Hamiltonian $H_{eff}$, which has less degrees of freedom, e.g., it has only $N$ eigenstates. The idea of the Monte Carlo Hamiltonian is that an effective Hamiltonian can be found via use of Monte Carlo, such that transition amplitudes become a finite sum over $N$ eigenstates, where $N$ is in the order of magnitude of $N_{c}$, i.e. the number of equilibrium configurations, sufficient to closely approximate the path integral of Eq.(\ref{eq:PathIntegral}).

\bigskip

One might ask: What is the virtue of such a Hamiltonian?
A list of physics problems, where progress has been slow with 
conventional methods including standard lattice techniques, and where such a Hamiltonian might bring progress are the following topics: \\
- Non-perturbative computation of cross sections and decay amplitudes in many-body systems\cite{kn:Kroger}. \\
- Low-lying but excited states of the hadronic spectrum 
and the related question of quantum chaos in such a system. \\
- Hadron wave functions and the related question of 
hadron structure functions, in particular for small $x_{B}$ and $Q^{2}$. 
The Hamiltonian formulation is suited to compute wave functions, which is quite difficult in the Lagrangian lattice formulation. \\
- Finite temperature and in particular finite density in baryonic matter. This is crucial for the quark-gluon plasma phase transition, the physics of neutron stars and cosmology.
The Hamiltonian formulation is suited to compute the mean value of the energy (average energy). This is difficult to compute in the Lagrangian lattice formulation where one usually computes the expectation value of the action. Finite density $QED$ and $QCD$ 
in the Lagrangian lattice formulation is hampered by the notorious complex action problem. \\
- Atomic physics: study of spectra and the question of quantum chaos. \\
- Condensed matter physics: study of spin systems (computation of dynamical structure factors), and high $T_{c}$ superconductivity models (search for 
electron pair attraction at very small energy).  
In the following we will outline how to construct such a Hamiltonian.

\bigskip

\section{Construction of $H_{eff}$}
In contrast to the statistical mechanics concept of the transfer matrix, which describes the time-evolution (we consider imaginary time) when advancing the system by a small discrete time step $\Delta t =a_{0}$ and from which one can infer the Hamiltonian $(a_{0} \to 0)$, here we consider transition amplitudes $<\psi | e^{-H T/\hbar} | \phi >$ corresponding to a finite, long time $T$ ($T >> a_{0}$),
for the purpose to reconstruct the spectrum in some finite low energy domain. Let us start from a complete orthonormal basis of Hilbert states $| e_{i}>, ~ i=1,2,3 \cdots $ and consider the matrix elements for a given fixed $N$
\be
M_{ij}(T) = <e_{i} | e^{-H T /\hbar} | e_{j} >, ~~~ i,j \in 1,\cdots,N .
\ee
Under the assumption that $H$ is Hermitian, $M(T)$ is a positive, Hermitian matrix. Elementary linear algebra implies that there is a unitary matrix $U$ and a real, diagonal matrix $D$ such that
\be 
M(T) = U^{\dagger} ~ D(T) ~ U .
\ee
On the other hand, projecting $H$ onto the the subspace $S_{N}$ generated by the first $N$ states of the basis $|e_{i}>$, and using the eigenrepresentation of such Hamiltonian, one has
\be
M_{ij}(T) = 
\sum_{k=1}^{N} < e_{i} | E^{eff}_{k} > e^{-E^{eff}_{k} T/\hbar} < E^{eff}_{k} | e_{j} > ,
\ee
and we can identify
\be
U^{\dagger}_{ik} = < e_{i} | E^{eff}_{k} >, ~~~ D_{k}(T) = e^{-E^{eff}_{k} T/\hbar } .
\label{eq:DefEnergyWaveFct}
\ee

\bigskip

Let us assume for the moment that the matrix elements 
$M_{ij}(T), ~ i,j = 1, \cdots, N$ would be known. 
Then algebraic diagonalization of the matrix $M(T)$ yields eigenvalues 
$D_{k}(T), ~ k=1, \cdots, N$, which by Eq.(\ref{eq:DefEnergyWaveFct}) 
gives the spectrum of energies,
\be
E^{eff}_{k} = - \frac{\hbar}{T} \ln D_{k}(T), ~~ k= 1, \cdots, N .
\ee
The corresponding k-th eigenvector can be identified with the 
k-th column of the matrix $U^{\dagger}_{ik}$. From Eq.(\ref{eq:DefEnergyWaveFct}) we then know the wave function of the k-th eigenstate expressed in terms of the basis $| e_{i}>$. 
Thus starting from the matrix elements $M_{ij}(T)$ we have explicitly constructed an effective Hamiltonian
\be
H_{eff} = \sum_{k =1}^{N} | E^{eff}_{k} > E^{eff}_{k} < E^{eff}_{k} | .
\ee

\section{Computation of matrix elements by Monte Carlo}
We suggest to compute the matrix elements $M_{ij}(T)$ directly from the action
via Monte Carlo with importance sampling. For the sake of simplicity, let us consider $D=1$. We choose basis states $|e_{i}>$ in position space by introducing a lattice with nodes $x_{i}$ and define $e_{i}(x)$ (unnormalized) 
by $e_{i}(x)=1$ if $x_{i} \leq x \leq x_{i+1}$, zero else. 
$\Delta x_{i}=x_{i+1}-x_{i}$. In numerical calculations we have used a regular lattice, $\Delta x_{i} = \mbox{const}$. The matrix elements read
\bea
M_{ij}(T) &=&
\int_{x_{i}}^{x_{i+1}} d y 
\int_{x_{j}}^{x_{j+1}} d z
< y, T | z, 0 >
\nonumber \\
&=& \left.  
\int_{x_{i}}^{x_{i+1}} d y 
\int_{x_{j}}^{x_{j+1}} d z
\int [dx] \exp[ - S[x]/\hbar ] \right|^{y,T}_{z,0} .
\eea
Here $S$ denotes the Euclidean action for a given path $C$,
\be
S[C] = \left. \int_{0}^{T} dt ~ \frac{1}{2} m \dot{x}^{2} + V(x) \right|_{C} .
\ee
The Monte Carlo method with importance sampling is suited and conventionally applied to estimate a ratio of integrals, like in Eq.(\ref{eq:PathIntegral}). 
Here we suggest to estimate the matrix elements $M_{ij}$ by splitting the action
\be
S = S_{0} + S_{V} \equiv 
\int_{0}^{T} dt ~ \frac{1}{2} m \dot{x}^{2} + 
\int_{0}^{T} dt ~ V(x) ,
\ee
and to express $M_{ij}$ as
\be
M_{ij}(T) = M^{(0)}_{ij}(T) ~
\frac{ 
\left.
\int_{x_{i}}^{x_{i+1}} d y 
\int_{x_{j}}^{x_{j+1}} d z
\int [dx] ~ \exp[ - S_{V}[x]/\hbar ] ~ \exp[ -S_{0}[x]/\hbar ] \right|^{y,T}_{z,0} }
{ \left.
\int_{x_{i}}^{x_{i+1}} d y 
\int_{x_{j}}^{x_{j+1}} d z
\int [dx] ~ \exp[ -S_{0}[x]/\hbar ] \right|^{y,T}_{z,0} } ,
\ee
where $O \equiv \exp[ -S_{V}/\hbar]$ is treated as an observable.
The ratio can be treated by standard Monte Carlo methods with importance sampling.
The matrix elements $M^{(0)}_{ij}$, corresponding to the free action $S_{0}$, 
are almost known analytically, 
\be
M^{(0)}_{ij}(T) =  
\int_{x_{i}}^{x_{i+1}} d y 
\int_{x_{j}}^{x_{j+1}} d z ~
\sqrt{ \frac{ m }{ 2 \pi \hbar T } } ~
\exp \left[ - \frac{ m}{2 \hbar T } (y - z)^{2} \right] .
\ee

\section{Test of $H_{eff}$}
\subsection{Free system}
In order to test the effective Hamiltonian, we have computed the energy spectrum, its wave functions and thermodynamic observables like the average energy $U$ and the specific heat $C$ as well as the partition function $Z$. They are defined by
\bea 
Z(\beta) &=& Tr[ e^{-\beta H } ] ,
\nonumber \\
U(\beta) &=& \frac{1}{Z} Tr[ H e^{ -\beta H } ] 
= - \frac{ \partial \log Z }{ \partial \beta } ,
\nonumber \\
C(\beta) &=& \frac{ \partial F }{\partial {\cal T}} 
= k_{B} \beta^{2} \frac{ \partial^{2} \log Z } { \partial \beta^{2} } ,
\label{eq:DefThermo}
\eea
where $\beta = (k_{B} {\cal T})^{-1}$, ${\cal T}$ is the temperature, and we identify $\beta$ with the imaginary time $T$ by $\beta = T/\hbar$.
For the free system one obtains the following analytical expressions for $Z$, $U$ and $C$,
\bea
Z(\beta) &=& \sqrt{ \frac{ m }{ 2 \pi \hbar^{2} \beta } } ~ I, ~~ 
I = \int_{-\infty}^{\infty} dx ~~ (\mbox{being infinite}) ,
\nonumber \\  
U(\beta) &=& \frac{1}{2 \beta} = \frac{1}{2} k_{B} {\cal T} ,
\nonumber \\
C(\beta) &=& \frac{1}{2} k_{B} .
\eea
Note that $U(\beta) \longrightarrow_{\beta \to \infty} 0$, i.e. it tends to the ground state energy of the free system (Feynman-Kac formula).

\bigskip

The partition function corresponding to the effective Hamiltonian is obtained via its spectrum,
\be
Z_{eff}(\beta) = Tr[ e^{- \beta H_{eff} } ] = 
\sum_{k=1}^{N} e^{-\beta E^{eff}_{k}} .
\ee
Via Eq.(\ref{eq:DefThermo}) one obtains the corresponding average energy $U_{eff}$
and the specific heat $C_{eff}$. One should keep in mind that $H_{eff}$ has been constructed for a specific value of the time parameter, $T=1$ corresponding to the temperature ${\cal T} =1$ (we use $\hbar = k_{B}=1$). Fig.[1] shows a plot of the average energy, comparing 
the exact result with that from the effective Hamiltonian. One observes that the agreement is better where ${\cal T} \to 0$, i.e. in the low energy regime.
A similar behavior is found for the specific heat, shown in Fig.[2].

\subsection{Harmonic oscillator}
The Euclidean action of the harmonic oscillator is given by 
\be
S = \int_{o}^{T} dt \frac{1}{2} m \dot{x}^{2} + \frac{1}{2} m \omega^{2} x^{2} .  \ee
The energy spectrum is
\be
E_{n} = \hbar \omega (n + 1/2), ~~ n=0,1,2,\cdots 
\ee
A comparison of the spectrum of the effective Hamiltonian with the exact one is 
shown in Tab.[1]. As can be seen, the error is small in the low energy domain. 
A more stringent test is that of the wave functions. Fig.[3] shows a comparison for the wave functions of the three lowest states.
We have also verified the low energy behavior of the effective Hamiltonian
by computing the partition function, average energy and specific heat as a function of temperature.
For the harmonic oscillator those are analytically known,
\bea
Z(\beta) &=& \left[ 2 ~ \sinh (\beta \hbar \omega/ 2) \right]^{-1} ~ ,
\nonumber \\
U(\beta) &=& \frac{\hbar \omega }{2} ~ \mbox{ctgh} (\beta \hbar \omega / 2) ~ ,
\nonumber \\
C(\beta) &=& k_{B} \left[ \frac{\beta \hbar \omega/2 } { \sinh (\beta \hbar \omega / 2) } \right]^{2} ~ .
\eea
In the limit $\beta \to \infty$ the average energy tends to 
the ground state energy, $U \to \hbar \omega/2$ (Feynman-Kac formula).
A plot of the average energy and the specific heat is shown in Figs.[4,5].
The effective Hamiltonian, constructed at 
$T_{c}=\beta_{c}={\cal T}_{c}=1$, 
describes well thermodynamic observables in the range 
$\beta_{c} \leq \beta$ ~ (it works also for $\beta > 10$, not shown in the figure). 
However, it breaks down for $\beta < \beta_{c}$, i.e. ${\cal T} > {\cal T}_{c}$.
This is due to the small dimension $N=20$ of the matrix. Agreement in a larger $\beta$-region, i.e. lowering $\beta_{c}$ can be obtained by increasing $N$. This can be seen, e.g. for the free system in Fig.[2], where $N=200$ and $\beta_{c} < 0.1$.

\subsection{Other local potentials}
We have tested the effective Hamiltonian for other local potentials. For example, 
\be
V(x) = - V_{0} ~ \mbox{sech}^{2}(x/d)
\label{eq:PotSech}
\ee
is a potential having a minimum $-V_{0}$ at $x=0$ 
and rising asymptotically to zero at $x=\pm \infty$.
It generates a bound state spectrum, being analytically known
\cite{kn:Morse}. It is given by
\bea
E_{n} &=& \frac{\hbar^{2}}{2 m d^{2} } \lambda_{n} ~ ,
\nonumber \\
\lambda_{n} &=& - \left[ (n + 1/2) - \sqrt{Q + 1/4} \right]^{2} ~,
\nonumber \\
Q &=& \frac{2 m  d^{2}}{\hbar^{2}} ~ V_{0},
\nonumber \\
n &=& 0,1,2,\cdots,n_{max} < \sqrt{Q +1/4} -1/2 .
\eea
The results are shown in Tabs.[2,3].

\section{Conclusion}
We have proposed to construct an effective low-energy Hamiltonian from the action via use of the Monte Carlo method. We have shown that the method works for a number of systems in $1-D$ quantum mechanics, by computing 
the spectrum, wave functions and thermodynamical observables. \\
- We have not given an error estimate of the statistical errors. The reason is that the statistical error of the matrix elements can be estimated easily, however, to get from that an error estimate of the energy spectrum is difficult.
We defer that to a later study. \\
- We have not discussed an application to a field theory or a many-body system, although this is the area where the method should prove to be most useful. The reason is that this requires a new step, namely a stochastic (Monte Carlo) selection from the set of basis functions. This is presently under investigation. \\
- In our opinion an effective low-energy Hamiltonian will be very useful 
in condensed matter physics, atomic physics, nuclear physics, and 
high energy particle physics.

\vskip 1cm

\noindent {\bf Acknowledgments} \\
\noindent H.K. would like to acknowledge helpful discussions with M. Creutz, 
B. Berg, W. Janke. and P. Amiot. H.K. and K.J.M.M. are grateful 
for support by NSERC Canada.
X.Q.L. is supported by the
National Natural Science Fund for Distinguished Young Scholars,
supplemented by the
National Natural Science Foundation of China, 
fund for international cooperation and exchange,
the Ministry of Education, 
and the Hong Kong Foundation of
the Zhongshan University Advanced Research Center.

\newpage
\begin{flushleft}
{\bf Figure Caption}
\end{flushleft}
\begin{description}
\item[{Fig.1}]
Average energy of the free system. 
Solid line and diamonds, respectively,
represent the exact analytical result, 
and that from the exact matrix elements for $\Delta x=0.5$ 
and $N=100$. The cross at $\beta=0.1$ corresponds to
$\Delta x=0.2$ and $N=200$.
\label{figF_free}
\item[{Fig.2}]
Specific heat over $k_B$ of the
free system. Symbols as in Fig.[1].
\label{figC_free}
\item[{Fig.3}]
Wave function of the harmonic 
oscillator, (a) ground state, (b) first excited state, 
(c) second excited state. 
Solid line, diamonds and crosses, 
respectively, represent the exact analytical result, 
that from the exact matrix elements,
and that from Monte Carlo simulation.
\label{figwave}
\item[{Fig.4}]
Average energy of the harmonic 
oscillator. Symbols as in Fig.[3].
\label{figF_harm}
\item[{Fig.5}]
Specific heat over $k_B$ of the harmonic 
oscillator. Symbols as in Fig.[3]. 
\label{figC_harm}
\item[{Tab.1}]
Eigenvalues of the harmonic
oscillator. $m=1$, $\hbar=1$, $\omega=0.6$,
$\Delta x=1$, $N=20$. $E^{exact}_{n}$, $E^{e.m.}_{n}$ and
$E^{m.c.}_{n}$, respectively, represent the exact analytical result, 
that from the exact matrix elements,
and that from Monte Carlo simulation.
\label{tab1}
\item[{Tab.2}]
Bound state spectrum for potential given by Eq.(\ref{eq:PotSech}).
(a) For $m=1.0, ~ \hbar=1.0, ~ T=1.0, ~ V_{0}=1.0, ~ d=1.0, ~ Q=2, 
~ \Delta x=1.0, ~ N=10$, there is only one bound state $n_{max} < 1$.
This is confirmed by the Monte Carlo data. 
(b) For $m=1.0, ~ \hbar =1.0, ~ T=1.0, ~ V_{0}=1.0, ~ d=2.0, ~ Q=8, 
~ \Delta x=1.0, ~ N=20$, there are three bound states $n_{max} < 3$.
This is confirmed by the Monte Carlo data.
\label{tab2}

\end{description}

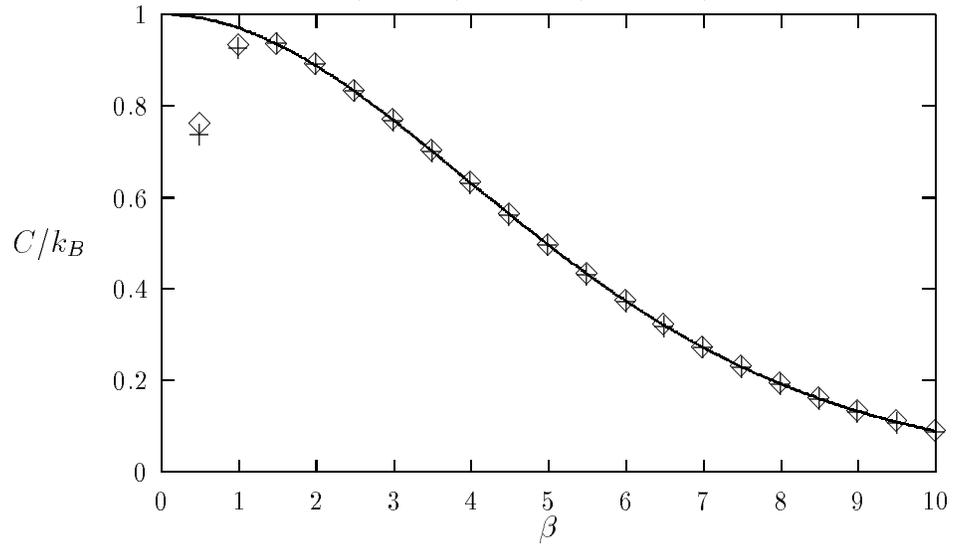
\end{document}